\documentstyle[aps,prl,epsf]{revtex}
\draft
\begin{document}
\twocolumn[\hsize\textwidth\columnwidth\hsize\csname
@twocolumnfalse\endcsname
\draft\pagenumbering{roma}
\author{ Yu-Xi Liu,C.P Sun $^{a,b}$, S.X.Yu, and D.L.Zhou}
\address{ Institute of Theoretical Physics, Academia Sinica, 
P.O.Box 2735, Beijing 100080, China}
\title{ Approach to the semiconductor cavity QED in
high-Q regimes with q-deformed boson}
\maketitle
\thispagestyle{empty}
\vspace{15mm}
\begin{abstract}
The high density Frenkel exciton which interacts with a single mode microcavity 
field is dealed with in the framework of the q-deformed boson. It is shown that the q-defomation
of bosonic commutation relations is satisfied naturally by the exciton operators when
the low density limit is deviated.
 An analytical expression of the 
physical spectrum for the exciton is given by using of the dressed states of 
the cavity field and the exciton. We also give the numerical study and compare
the  theoretical results with the experimental results
\end{abstract}
\vspace{4mm}
\vskip2pc] 
\newpage \pagenumbering{arabic}

\section{\bf \ Introduction}

It's well known that the exciton system is a quasi-particle system. At the
low density, excitons are approximately treated as bosons which obey Bose
statistics~\cite{A,B,C}. But when the density of the excitons become higher,
the excitons, which somewhat deviate ideal bosons, are no longer ideal
bosons. There are two ways of dealing with this problem : one way is to put
these deviations into the effective interaction between the hypothetical
ideal bosons and the exciton operators are still presented by the bosonic
operators~\cite{D,F}. Another way is the implement of the atomic operators
model~\cite{G,H,I}. The question naturelly arises whether the exciton system
is equivalent to the atomic system.

In this paper we show that the exciton system could be described by $q-$
deformed boson system which interpolates between Bose system and Fermi system
\cite{q1} ; and the deformation parameter $q$ is well defined by the total
atomic particle number $N,$rather than it is phenomenological in the
previous discussions .The concept of the q-deformed boson was even
extensively applied due to the q-deformed boson realization of quantum group
theory by different authors ten years ago~\cite{q2,q3,q4,q5}. Since then,
many physicists make efforts to find its real physical realizations. For
example, they give some phenomenological investigations to fit the deformed
spectra of rotation and oscillation for molecules and nuclei~\cite
{bo,chang,li}. In our opinion, those investigations can be regarded as
merely phenomenological because a $q$-deformed structure is postulated in
advance without giving it a microscopic mechanism. In this paper it will be
shown that a physical and natural realization of the $q$-deformed boson is
provided by the exciton operators, which was presented recently by
Gardiner \cite{Gardiner} for the description of Bose-Einstein condensation
(BEC). In fact, the similar quasiparticles scheme for particle-number
conservation has already been introduced by Girardeau and Arnowitt almost 40
years ago \cite{g1}. The relationship between Gardiner's phonons and these
quasiparticles has been discussed in a recent comment \cite{g2} .

Following these ways, we find that the exciton system also could be
described by the q-deformed boson. When the density of the exciton (the
particles excited in upper state) is low enough, it return to the ideal
boson case. Using this theory, we could give a good explanation on the
semiconductor cavity QED in high-Q regimes.What will be investigated here is
the case that the total atomic particle number $N$ is very large but not
infinite. That is, we shall consider the effects of order $o(1/N)$. And we
shall focus on an algebraic method, a q-deformed boson algebra, of treating
the effects of finite  particle number. As it turns out, the
commutation relations for the exciton operators will no longer obey the
commutation relation of the Heisenberg-Weyl algebra but the $q$-deformed
bosonic commutation relation 
\begin{equation}
\left[ b_q,b_q^{\dagger }\right] _q\equiv b_qb_q^{\dagger }-qb_q^{\dagger
}b_q=1,  \label{1}
\end{equation}
where the deformation constant $q$ depends on the total atomic particle
number.

This paper is organized as follows. In section 2 we firstly deduce the
q-deformed commutation relation for the exciton in the high-Q cavity in case
of the large but finite lattice molecule number $N$ . In section 3,  only
keeping  the first order term of $\frac 1N,$we model the Frenkel excitons in
a micro-cavity as the dressed q-deformed boson system. In section 4, the
quantum  approach for angular momentum  is used to obtain the eigen- values
and eigen-function of the system under the one order approximation. The
stationary physical spectrum of the system is calculated  in the section 5.
Finally we summarize   our results with some comments.

\section{$q$-deformed bosonic algebra for exciton}

Gardiner's starting point \cite{Gardiner}to introduce the exciton operators
is to consider a system of the weakly interacting Bose gas. Without losing
generality, we consider a thin molecular crystal film containing $N$
identical two-level molecules interacting resonantly with a single mode
quantum field. The intermolecular interaction is neglected. We assume that
all molecules have equivalent mode positions, so they have the same coupling
constant $\kappa$. By using Dick model~\cite{a}, we could write the Hamiltonian
under the rotating wave approximation as following: 
\begin{equation}
H=\hbar \Omega (S_z+a^{+}a)+\hbar \kappa (aS_{+}+a^{+}S_{-}),
\end{equation}
where, $a$ is annihilation operator of the quantum cavity field and 
\begin{equation}
\left. 
\begin{array}{lll}
S_Z & = & \sum_{n=1}^Ns_z(n), \\ 
S_{+} & = & \sum_{n=1}^Ns_{+}(n), \\ 
S_{-} & = & \sum_{n=1}^Ns_{-}(n),
\end{array}
\right. 
\end{equation}
where, $s_z(n)=\frac 12(|e_n><e_n|-|g_n><g_n|)$, $s_{+}(n)=|e_n><g_n|$ and $%
s_{-}(n)=|g_n><e_n|$ are quasi spin operators of the nth molecule. $|e_n>$
and $|g_n>$ are the excited state and the ground state of n'th molecule.

Consider the second quantization of the above model. Let $b_e^{\dagger }$
and $b_e$ denote the creation and annihilation operators for the atoms in
the excited state and $b_g^{\dagger }$ and $b_g$ for the creation and
annihilation operators of the atoms in the ground state. The simplified
Hamiltonian in second quantization reads 
\begin{equation}
H=\hbar \Omega (b_e^{\dagger }b_e-b_g^{\dagger }b_g+a^{+}a)+\hbar \kappa
[ab_e^{\dagger }b_g+H{\rm .c.]}.  \label{h1}
\end{equation}
Note that the total atomic particle number ${\bf N}=b_e^{\dagger}b_e
+b_g^{\dagger }b_g$ is conserved. For convenience we define $\eta =1/{N}$
for large particle number.

In the thermodynamical limit ${N}\to \infty $, the Bogoliubov approximation 
\cite{bogo,book}is usually applied, in which the ladder operators $%
b_g^{\dagger },b_g$ of the ground state are replaced by a $c$-number $\sqrt{%
N_c}$, where $N_c$ is the average number of the ininital condensated atoms.
As a result Hamiltonian Eq.(\ref{h1}) becomes a two-coupling harmonic
oscillator system

\begin{equation}
H_b=\hbar \Omega (b_e^{\dagger }b_e+a^{+}a)+\hbar \kappa \sqrt{N_c}%
[ab_e^{\dagger }+H{\rm .c.]}.
\end{equation}
However, this apporoximation destroyes a symmetry of the Hamiltonian Eq.(\ref
{h1}), i.e., the conservation of the total particle number is violated
because of $[{N},H_b]\neq 0$.

To avoide this problem, the exciton operators are defined as: 
\begin{equation}
b_q=\frac 1{\sqrt{{N}}}b_g^{\dagger }b_e,\qquad b_q^{\dagger }=\frac 1{\sqrt{%
{N}}}b_gb_e^{\dagger }.
\end{equation}
according to Gardiner\cite{Gardiner}.These operators act invariantly on the
subspace $V^N$ spanned by bases $|N;n\rangle \equiv |N-n,n\rangle $ $%
(n=0,1,\ldots ,N)$, where Fock sates $(m,n=0,1,2,\ldots )$ 
\[
|m,n\rangle =\frac 1{\sqrt{m!n!}}b_e^{\dagger m}b_g^{\dagger n}|0\rangle 
\]
spann the Fock space $H_{2b}$ of a two mode boson.

A straightforward calculation leads to the following commutation relation
between the exciton  operaotr and its Hermitian conjugate: 
\begin{equation}
\left[ b_q,b_q^{\dagger }\right] =1-\frac {2}{N}b_e^{\dagger}b_e
=f(b_q^{\dagger }b_q;\eta ),
\end{equation}
with $f(x;\eta )=\sqrt{1+2(1-2x)\eta +\eta ^2}-\eta$. Keeping only the
lowest order of $\eta $ for a very large total  particle number, the
commutator above becomes 
\begin{mathletters}
\begin{equation}
\left[ b_q,b_q^{\dagger }\right] =1-2\eta b_q^{\dagger }b_q
\end{equation}
or 
\begin{equation}
\left[ b_q,b_q^{\dagger }\right] _q=1,  \label{bcr}
\end{equation}
\end{mathletters}
with $q=1-2\eta $. This is exactly a typical $q$-deformed commutation
relation. As $N\to \infty $ or $q\to 1$, the usual commutation relation of
Heisenberg-Weyl algebra is regained.

In the above discussion about the phonon excitation, we have linearized
commutator $h\equiv f(b^{\dagger }b;\eta )$ so that a $q$-deformed
commutation rule was obtained. Essentially this linearization establishes a
physical realization of the $q$-deformed algebra. However, if the total
particle number $N$ is not large enough, then $h$ can not be approximated by
a linear function. From the commutation relations between $h$ and 
$b_q,b_q^{\dagger}$

\begin{equation}
\left[ h,b_q^{\dagger }\right] =-\frac{2}{N}b_q^{\dagger },\qquad [h,b_q]=
\frac{2}{N}b_q,
\end{equation}
we see that the algebra of exciton operators is a rescaling of algebra 
$su(2)$ with factor $N$.

\section{Theoretical Model}

Based on the above analysis about the algebraic structure of exciton
operator, we consider the case of the low density of atoms in excited sate
for the Hamiltonian (2).

Since the second quantization forms of $S_{+}$ and $S_{-}$ are 
$S_{+}=b_e^{\dagger }b_g,S_{-}=b_g^{\dagger }b_e$, it is straighforward to
prove that the collective operator $\frac{S_{+}}{\sqrt{N}}$ and 
$\frac{S_{-}}{\sqrt{N}}$ are approximately considered as the simple bosonic operators as 
$N\to \infty .$ These collective operators are called exciton operators. But
in case of the high density of molecules in excited state with finite $N$,
many molecules are in the excited states, the bosonic approximation could no
longer work well. The Hamiltonian (2) is rewritten as the effective 
Hamiltonian
in the form of q-deformed boson: 
\begin{equation}
H=\hbar \Omega (a^{+}a+b_q^{+}b_q)+\hbar g(a^{+}b_q+b_q^{+}a)
\end{equation}
with $g=\sqrt{N}k $, $b_q$ and $b_q^{+}$ satisfy q-deformed relation: 
\begin{equation}
\lbrack b_q,b_q^{+}]=b_qb_q^{+}-qb_q^{+}b_q=1,
\end{equation}
where,
\begin{equation}
q=1-\frac{2}{N}
\end{equation}
Here, the  deformation parameter $q$ is determined by the lattice 
molecule  number. So $q$ is no longer phenomenological.

Up to the first order approximation, $b_q^{+}(b_q)$ could be expressed as
following 
\begin{eqnarray}
b_q^{+} &=&b^{+}+\frac{b^{+}b^{+}b}{2N}, \\
b_q &=&b+\frac{b^{+}bb}{2N}.
\end{eqnarray}
in terms of the general bosonic operators $b^{+}(b)$ .So the Hamitonian $H$
is rewritten in form of perturbation 
\begin{equation}
H=H_0+H^{\prime }
\end{equation}
where 
\begin{eqnarray}
H_0 &=&\hbar \Omega (a^{+}a+b^{+}b)+\hbar g(a^{+}b+b^{+}a), \\
H^{\prime } &=&\frac \hbar {2N}(2\Omega
b^{+}b^{+}bb+gb^{+}b^{+}ab+a^{+}b^{+}bb).
\end{eqnarray}
It is clearly that $H^{\prime }$ is equivalent to the attractive
exciton-exciton collisions due to the bi-exciton effect and decreased
exciton-photon coupling constants due to the phase phase filling 
effect~\cite{chi}
.

\section{\bf Approximate analytical solutions}

To solve the Schroedinger equation governed by eq.(15),we implement the
quantum angular momentum theory \cite{bei}

If we define the angular momentum operators 
\begin{eqnarray}
J_z &=&\frac 12(a^{+}a-b^{+}b),  \nonumber \\
J_{+} &=&a^{+}b,J_{-}=ab^{+}.
\end{eqnarray}
then 
\begin{eqnarray}
J_{x} &=&\frac{1}{2}(a^{+}b+ab^{+}),  \nonumber \\
J_{y} &=&\frac{1}{2i}(a^{+}b-ab^{+}).
\end{eqnarray}
We rewrite the Hamitonian (5)  
\begin{equation}
H_0=\hbar \Omega \hat{N}+2\hbar gJ_x=\hbar\Omega\hat{N}+
2\hbar ge^{-i\frac \pi 2J_y}J_z e^{i\frac \pi 2J_y}.
\end{equation}
In terms of a $SO(3)$ rotation $\hbar (\Omega \hat{N}+2gJ_x)$ by  
$e^{i\frac \pi 2J_y}$.  Note that the
excitation number operator 
$\hat{N}=a^{+}a+b^{+}b$ is a constant under the a $SO(3)
$ rotation and 
\begin{equation}
J^2=J_x^2+J_y^2+J_z^2=\frac{\hat{N}}{2}(\frac{\hat{N}}{2}-1)
\end{equation}
is the total angular momentum operator. In terms of the eigen states of $J^2$
and $J_z$  
\begin{equation}
|jm>=\frac{(a^{+})^{j+m}(b^{+})^{jm}}{\sqrt{(j+m)!(j-m)!}}|0>,
\end{equation}
where the eigen values of the $J^2$ and $J_z$ are 
\begin{equation}
j=\frac {\cal N}{2},m=-\frac{\cal N}{2},\cdots ,\frac{\cal N}{2}
\end{equation}
the eigen functions and the eigen values of $H_0$ are constructed as  
\begin{equation}
\psi _{jm}^0=e^{-i\frac \pi 2J_y}|jm>,
E_{jm}^{(0)}=\hbar \Omega{\cal N}+2\hbar gm.
\end{equation}

Up to the first  order approximation, the eigen values of $H$ are obtained
as  
\begin{equation}
E_{jm}=E_{jm}^{(0)}+<mj||e^{i\frac \pi 2J_y}H^{\prime }e^{-i\frac \pi 2J_y}|jm>,
\end{equation}
whith their corresponding eigen functions  
\begin{equation}
\psi _{jk}=\psi _{jk}^{(0)}+\sum_{n\ne k}\frac{H_{nk}^{\prime}}
{E_{jk}^{(0)}-E_{jn}^{(0)}}\psi _{jn}^{(0)}.
\end{equation}
where,$n$ and $k$ present the subscript $(jm^{\prime })$ and $(jm)$. We
calculate the matrix elements of the pretubation Hamiltonian $H^{\prime}$:  
\begin{eqnarray}
&<&m^{\prime }j||e^{i\frac \pi 2J_y}H^{\prime }e^{-i\frac \pi 2J_y}|jm> 
\nonumber \\
&=&\frac \hbar {4N}\Omega \sqrt{(j+m)(j+m-1)}  \nonumber \\
&\times &\sqrt{(j-m+1)(j-m+2)}\delta _{m-2,m^{\prime }}  \nonumber \\
&+&\frac \hbar {4N}\Omega \sqrt{(j+m+1)(j+m+2)}  \nonumber \\
&\times &\sqrt{(j-m)(j-m-1)}\delta _{m+2,m^{\prime }}  \nonumber \\
&+&\frac \hbar {4N}(2\Omega -g)\sqrt{(j-m)(j+m+1)}  \nonumber \\
&\times &(j-m-1)\delta _{m+1,m^{\prime }}  \nonumber \\
&+&\frac \hbar {4N}(2\Omega +g)\sqrt{(j-m)(j+m+1)}  \nonumber \\
&\times &(j+m)\delta _{m+1,m^{\prime }}  \nonumber \\
&+&\frac \hbar {4N}(2\Omega +g)\sqrt{(j+m)(j-m+1)}  \nonumber \\
&\times &(j+m-1)\delta _{m-1,m^{\prime }}  \nonumber \\
&+&\frac \hbar N(2\Omega -g)\sqrt{(j+m)(j-m+1)}  \nonumber \\
&\times &(j-m)\delta _{m-1,m^{\prime }}  \nonumber \\
&+&\frac \hbar {4N}(\Omega +g)(j+m)(j+m-1)\delta _{m,m^{\prime }}  \nonumber
\\
&+&\frac \hbar {4N}(\Omega -g)(j-m)(j-m-1)\delta _{m,m^{\prime }}  \nonumber
\\
&+&\frac \hbar N\Omega (j^2-m^2)\delta _{m,m^{\prime }}
\end{eqnarray}
so the eigen values of $H$ are 
\begin{eqnarray}
E_{jm} &=&\hbar \Omega {\cal N}+2m\hbar g+\frac \hbar N\Omega (j^2-m^2)  \nonumber
\\
&+&\frac \hbar {4N}(\Omega +g)(j+m)(j+m-1)  \nonumber \\
&+&\frac \hbar {4N}(\Omega -g)(j-m)(j-m-1)
\end{eqnarray}
In general, we could obtain all eigen functions of $H$ under one order
approximation by using eq.(24), eq.(26), and eq.(27). So
the time evolution operator of the system is written as: 
\begin{equation}
U(t)=exp(-i\frac H\hbar t)=\sum_{j=0}\sum_{m=-j}^jexp(-i\frac{E_{jm}}\hbar
t)|\psi _{jm}><\psi _{jm}|
\end{equation}

\section{\bf Fluorescence spectrum of the exciton }

We firstly give an analytic expression for the physical spectrum of the
q-deformed exciton in terms of the Fock state of the quantum field and the
exciton. The standard definition of the physical spectrum is~\cite{b} 
\begin{equation}
S(\omega )=2\gamma \int_0^t{\rm d}t_1\int_0^t{\rm d}t_2e^{-(\gamma -i\omega
)(t-t_2)}e^{-(\gamma +i\omega )(t-t_1)}G(t_1,t_2)
\end{equation}
where $\gamma $ is the half-bandwith of spectrometer which is being used to
measure the spectrum, and $t$ is time length of the excitation in the
cavity. $G(t_1,t_2)$ is dipole correlation function, and 
\begin{equation}
G(t_1,t_2)=<i|U^{+}(t_2)b_q^{+}U(t_2)U^{+}(t_1)b_qU(t_1)|i>,
\end{equation}
where, $|i>$ is any initial state of the system. By using the bosonic
approximation  and substituting eq.(29) into eq.(31), the dipole
correlation function is rewritten as following: 
\begin{eqnarray}
&&G(t_1,t_2)=<i|U^{+}(t_2)b_q^{+}U(t_2)U^{+}(t_1)b_qU(t_1)|i>  \nonumber \\
&=&<i|U^{+}(t_2)(b^{+}+\frac{b^{+}b^{+}b}{2N})U(t_2)U^{+}(t_1)(b+
\frac{b^{+}bb}{2N}))U(t_1)|i>  \nonumber \\
&=&\sum_{j,k,l,m,n,}<i|\psi _{jl}><\psi _{jl}|(b^{+}+\frac{b^{+}b^{+}b}{2N}
)|\psi _{km}>  \nonumber \\
&&\times <\psi _{km}|(b+\frac{b^{+}bb}{2N}))|\psi _{jn}><\psi
_{jn}|i>e^{i\omega _{lm}t_2}e^{-i\omega _{nm}t_1}
\end{eqnarray}
where $\omega _{lm}=\frac{E_{jl}-E_{km}}\hbar $ and $\omega _{nm}=\frac{
E_{jn}-E_{km}}\hbar $. It's evident that $j$ is determined only by the
initial state $|i>$. So we have 
\begin{eqnarray}
S(\omega ) &=&\sum_{l,k,m}\frac{2\gamma }{\gamma ^2+(\omega -\omega _{lm})^2}
|<i|\psi _{jl}>|^2  \nonumber \\
| &<&\psi _{jl}|(b^{+}+\frac{b^{+}b^{+}b}{2N})|\psi _{km}>|^2
\end{eqnarray}
Noting that we have passed the transient terms and slowly variation terms.
This equation gives the stationary physical spectrum in terms of the system
eigenvalues and eigenstaes. If $<m^{\prime }j^{\prime }|b^{+}|jm>\not=0$,
then we have $j^{\prime }=j+\frac 12$ and $m^{\prime }=m-\frac 12$. So the
equation (25) is rewritten as: 
\begin{eqnarray}
S(\omega ) &=&\sum_{l,m}\frac{2\gamma }{\gamma ^2+(\omega -\omega _{lm})^2}
|<i|\psi _{jl}>|^2  \nonumber \\
| &<&\psi _{jl}|(b^{+}+\frac{b^{+}b^{+}b}{2N})|\psi _{(j-\frac 12)m}>|^2
\end{eqnarray}
The eigenvalues determine the position of the spectral component and $
|<i|\psi _{jl}>|^2|<\psi _{jl}|(b^{+}+\frac{b^{+}b^{+}b}{2N})|\psi _{(j-
\frac 12)m}>|^2$ determine the intensity of the spectral lines.

In terms of the experimental condition of the reference~\cite{bb}, the bare
excitons could be prepared by resonant femtosecond pulse pumping. If we
prepare the initial state being in ${\cal N}=1$, then the eq.(34) shows
 that
the emission spectrum of the ${\cal N}=1$ to the ${\cal N}=0$ transition has 
double peaks
structure which is exactly equal to that of the two-level atomic 
system.
When the pumping power is increased , the emission spectrum is quiet
different from the case of the two-level atomic system. For example, if the
system initially is in Fock state ${\cal N}=2$ , then we have: 
\begin{eqnarray}
S(\omega ) &=&\sum_{l,m}\frac{2\gamma }{\gamma ^2+(\omega -\omega _{lm})^2}
|<i|\psi _{1l}>|^2  \nonumber \\
| &<&\psi _{1l}|(b^{+}+\frac{b^{+}b^{+}b}{2N})|\psi _{\frac 12m}>|^2
\end{eqnarray}
From this equation we know when ${\cal N}=2$ there are six peaks in the emission
spectrum which are expected. Although there are three different initial
state, they have similar spectrum shape. As is shown in Fig.1, this sextet
structure is contrast to the triplet structure in the emission spectrum from
strong pumped two-level system~\cite{br}.

\begin{figure}
\epsfxsize=6 cm
\centerline{\epsffile{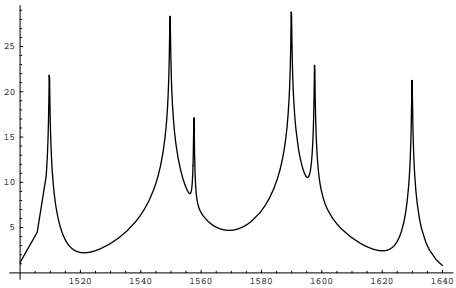}}
\caption[]{$\Omega$=1562 mev, N=100, g=20 mev, $\gamma$=0.1 mev}
\end{figure}
When the molecular number of the
system is increased, Such as there are 10000 molecular in the system, the 
other conditions are the same as that of the Fig.1, 
then the coupling between the molecular and the cavity field is weak. There 
are two peaks in the emission spectrum
(Fig. 2).
In this case Bose 
approximation is good.
\begin{figure}
\epsfxsize=6 cmq
\centerline{\epsffile{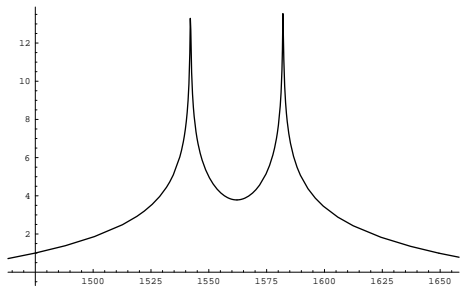}}
\caption[]{$\Omega$=1562 mev, N=10000, g=20 mev, $\gamma$=0.1 mev}
\end{figure}

\section{\bf Conclusion}
It has been shown that the higher density Frenkel exciton  coupling to  a
single mode   high-Q  microcavity field can be described by the quantum
dynamics for the  dressed  q-deformed  boson. Keeping the first order term
of $\frac 1N,$the high density Frenkel exciton naturally obyes the
q-deformed commutation relation. Based on this observation the quantum
theory of  angular momentum is used to obtain the eigen- values and
eigen-function of the exciton system under the one order approximation.
Comparing with the usual  approach for  Frenkel exciton dynamics  our 
Hamiltonian is Hermitian and closed in form. An analytical expression for
the stationaryphysical spectrum for the exciton is obtained by using of the
dressed states of the cavity field and the exciton. 

\begin{center}
{\bf Acknowledgment}
\end{center}

{\it This work is  supported by the NSF of China and ''9-5 Pandeng ''
project.}

\end{document}